\documentstyle[aps,preprint]{revtex}

\begin{document}

\title { $1/N_c$- expansion of the quark condensate at finite temperature }
\author{ D. Blaschke,
        Yu.L. Kalinovsky
        \thanks{Supported by DFG  Grant No. Ro 905/7-2 },
 G. R\"opke, S. Schmidt and M.K. Volkov
\thanks{Permanent address: Bogolubov Laboratory of Theoretical Physics,
        JINR Dubna, Russia} }

\address{MPG Arbeitsgruppe {"Theoretische
          Vielteilchenphysik"} \\
Universit\"at Rostock, D-18051 Rostock, Germany}

\maketitle

\begin{abstract}
Previously the quark and meson properties in a many quark system at finite
temperature have been studied within effective QCD approaches in the
Hartree approximation.
In the present paper we consider the influence of the mesonic correlations on
the quark self-energy and on the quark propagator within a systematic
$1/N_c$- expansion. Using a general separable ansatz for the nonlocal 
interaction, we derive a selfconsistent equation for the $1/N_c$ correction to
the quark propagator. For a separable model with cut-off formfactor, we obtain 
a decrease of the condensate of the order of 20\% at zero temperature.
A lowering the critical temperature for the onset of the chiral restoration
transition  due to the inclusion of mesonic correlations is obtained what 
seems to be closer to the results from lattice calculations. 

\vspace{0.4cm}

\end{abstract}
\vspace{0.4cm}

\section{Introduction}
QCD motivated effective theories are the most promising approaches to the low 
energy behaviour of QCD and the meson physics in terms of quark and gluon
degrees of freedom and symmetries. Starting from chiral quark model 
Lagrangians 
a perturbative approach to the occurence of a chiral condensate below a
critical
temperature $T_c$  in mean field approximation is usually considered. 
Simultaneously, the pseudoscalar Goldstone boson, the pion, occurs. 
Perturbation theory can be formulated in $1/N_c$, where $N_c$ is the number of 
colors \cite{thooft}.
The leading order is the Hartree approximation, results are reported in  
Refs.\cite{nJl,njl,NJL,hatsuda}.
A more general approach, where a nonlocal instantaneous interaction is 
applied, 
has been presented in Refs.
\cite{itobuck,buballa,birse,short,physref,kaliprep}. 
A still open but very important question is the influence of mesonic degrees of 
freedom which are neglected in the Hartree approximation.
These degrees of freedom are supposed to be dominant in the low temperature 
limit.
For the NJL model, an effective $1/N_c$- expansion which accounts for
the mesonic fluctuations has been considered in \cite{sandi}.
However, the set of diagrams for the self-energy in next to leading order 
considered in this reference was not complete.
This has been observed in Ref. \cite{volkov} where also the r\^{o}le of the
scalar iso-vector mesons in the $1/N_c$ - approximation was discussed at zero 
temperature.
It was shown that  in the $1/N_c$- expansion the Schwinger Dyson equation 
for the quark self-energy is different from  the gap equation for the quark 
condensate and has to be solved separately.
A complete collection of diagrams in $1/N_c$ was given in Ref. \cite{snyderman}
and recently studied in the chiral limit $m_0=0$ at $T=0$ by Ref.\cite{lemmer}.
At $T=0$, effects of the order of $10\%-20\%$ have been obtained
in these approaches, showing that mesonic fluctuations play an important role.

In this work, we consider the influence of mesonic correlations
on the quark condensate at finite temperature. It is expected
that such a  calculation beyond the Hartree level of description will lead to
corrections to the temperature behaviour of the quark condensate since
the medium allows for mesonic degrees of freedom.
The relation of a generalized gap equation to the thermodynamical potential of 
a quark meson plasma has been considered in Ref. \cite{zhuang}.
The present paper is a first step for a consistent description of a meson gas at 
finite temperature within a chiral quark model.

The paper is organized as follows: In Section \ref{sec:model} the nonlocal 
chiral quark model is briefly introduced, which is used in Section 
\ref{sec:expan} to derive a generalized formula  for the quark condensate in  
${\cal O} (1/N_c)$ expansion.
In Section \ref{sec:meson} we include dynamical fluctuations into the 
self-energy and treat the scalar und pseudoscalar contributions within the
 pole 
approximation.
The numerical results for a calculation within the NJL model at 
finite temperature are discussed in Section \ref{sec:disc}.

\section{The model}
\label{sec:model}
Our starting point is the chiral symmetric effective Lagrangian in the
quark sector of the following general form
\begin{equation} 
\label{lag}
{\cal L} ={\bar q_1}(p)(\gamma_\mu p^\mu - m_0)q_1(p) + {\cal L}_{\rm int}~~,
\end{equation} 
where the interaction term
\begin{equation} 
{\cal L}_{\rm int} = -\frac{1}{2} 
{\bar q}_1(p_1)\Lambda_{12}^\phi  q_2(p_2)
K  (p_1,p_2,p_{1'},p_{2'})
{\bar q}_{2'}(p_{2'})\Lambda_{1'2'}^\phi  q_{1'}(p_{1'})
\end{equation} 
is given as a nonlocal generalization of the current-current type  
interaction. 
Here the matrices $\Lambda^\phi _{12}$ denote the decomposition into the
color (c), flavor (f) and Dirac (D) channels. 
In this work we restrict us to scalar and pseudoscalar channels. 
Therefore we choose 
$\Lambda^\sigma_{12} = [1_c\cdot 1_f\cdot 1_D]_{12}$ and 
$\Lambda^\pi_{12} = [1_c\cdot {\bf \tau}_f\cdot i\gamma_5]_{12}$.

The gluonic degrees of freedom do not occur explicitely in this effective 
approach to the low-energy sector of QCD. They are assumed to form a
 condensate 
which is responsible for the nonperturbative character of the quark- quark 
interaction in this domain.  
We make the phenomenological ansatz of an instantaneous interaction kernel, 
which can be formulated in a covariant way \cite{kaliprep}.
We employ here a separable form for the nonlocal 4-point 
interaction
of the form
\begin{equation} 
\label{sep}
K(p_1,p_2,p_1',p_2') = - \frac{K_0}{N_c} g(\frac{|{\bf p_1+p_2}|}{2})
 g(\frac{|{\bf p_1'+p_2'}|}{2}) 
\delta_{p_1-p_2,p_1'-p_2'}~.
\end{equation} 
The $N_c$- dependence  arises from the Fierz transformation of the quark
current-current interaction in the colour singlet channel considered here, 
see e.g. \cite{erv94}. 
For our numerical calculations in Section \ref{sec:disc} we use $N_c=3$.
The dependence of the formfactor on the modulus of the three-momentum 
($|{\bf p}|={\rm p} $) has been discussed for different shapes, e.g. 
a Gaussian one
($g({\rm p} )=\exp[-({\rm p} /\Lambda_{\rm Gauss})^2]$) or 
the well-known NJL type interaction
($ g({\rm p} )= \Theta(1-{\rm p} /\Lambda_{\rm NJL})$), see \cite{physref}. 
Note that the potential does not depend on the energy and we obtain therefore 
the NJL- model with a three-momentum cut--off.
The  spectral properties of the quark model  defined by the Lagrangian 
(\ref{lag}) are obtained from the single particle propagator
\begin{equation} 
G_{12} (p_1p_2) = [G(p_1)1_c1_f]_{12}\delta_{p_1,p_2}~,
\end{equation} 
which is a diagonal matrix in color, flavor and momentum space. 
The matrix element $G(p)$ obeys the Dyson equation
\begin{equation} 
\label{dyson}
G(p) =[G_0^{-1}(p)-\Sigma (p)]^{-1}~~ ,
\end{equation} 
where $G_0^{-1}(p)=\gamma_\mu p^\mu - m_0$ is the vacuum Green function, 
see Fig. 1.
The self-energy $\Sigma(p)$ is defined  by  an analysis of all one particle 
irreducible diagrams contributing to the propagator.
Having the single particle propagator at our disposal, the physical quantity
 of 
interest which is straightforwardly evaluated is the quark condensate.
For our separable potential we introduce the nonlocal quark condensate as
\begin{equation} 
\label{condensate}
<{\bar q} q>=  N_cN_f\sum_pg({\rm p} ){\bf Tr}\left[G(p)\right]~,
\end{equation} 
where ${\bf Tr}$ stands for the trace over the Dirac space only.
The  finite temperature investigations are performed using the Matsubara 
technique 
\cite{kapusta,fetter,kker},
where $p_0=i \omega_n$ with the fermionic Matsubara frequencies 
$\omega_n=(2n+1)\pi T$ and $\sum_p$  stands short for  
$T\sum_n\int {d{\bf p}}/{(2\pi)^3}$.
In order to obtain estimates for the quark condensate one has to make 
approximations for the self-energy.

The first step towards a systematic investigation of the Dyson
equation (\ref{dyson}) is the selfconsistent Hartree approximation, see Fig. 2,
\begin{equation} 
\label{selfhartree}
\Sigma^H [{\rm p} ;G^H] = -K_0 N_f g({\rm p} )\sum_k g({\rm k} )
{\bf Tr} \big[G^H(k)\big ] ~~,
\end{equation} 
which defines upon insertion in (\ref{dyson}) the propagator in Hartree 
approximation
\begin{equation}
\label{gh} 
G^H(k) = \left[G_0^{-1}(k) - \Sigma^H[{\rm k} ;G^H]\right]^{-1}~~.
\end{equation} 
The Hartree self-energy (\ref{selfhartree}) is a Dirac scalar and appears as a 
mass term in the propagator,
\begin{equation} 
\label{mh}
m^H({\rm p} )=m_0 - g({\rm p} ) \frac{K_0}{N_c} <{\bar q}q>^H
\end{equation} 
with a momentum dependence due to the nonlocality of the interaction kernel 
(\ref{sep}).
The quark condensate in Hartree approximation is, cf. Eqs. (\ref{gh}) - (\ref{condensate}),
\begin{eqnarray}
\label{condTH} 
<{\bar q}q>^H &=& N_cN_f\sum_k g({\rm k} ){\bf Tr}\bigg[ G^H(k)\bigg]
\nonumber\\
&=& -2N_cN_f\int\frac{d^3{\bf k}}{(2\pi)^3}g({\rm k} )\frac{m^H({\rm k} )}
{E({{\rm k}  })}[1-2f(E({{\rm k} }))]~,
\end{eqnarray} 
for details see e.g. Refs. \cite{nJl,njl,NJL,hatsuda,physref}.
In this approximation,  the magnitude as well as the temperature  dependence
of the  dynamical mass generation is determined from the condensate only.
Note that the restoration of the chiral symmetry at temperatures above the
critical  one ($T_c\sim 200 $ MeV, \cite{physref}) is governed by the Fermi 
distribution function of quarks in the medium,
$f (E({\rm k} )) = \left\{\mbox{exp}
\left[ \left( E({\rm k} ) \right)/T   \right]+1  \right\}^{-1}$,
where the quasiparticle dispersion relation
\begin{equation} 
E({\rm k} )=\sqrt{{\rm k} ^2+(m^H({\rm k} ))^2}
\end{equation} 
contains the momentum dependent Hartree mass (\ref{mh}).

It is, however, questionable whether the Hartree approximation is appropriate 
for the description of the nonperturbative  low energy region of QCD where
 free 
quarks should be absent due to confinement. 
Since  mesonic correlations are supposed to dominate the low energy excitation 
spectrum of the quark matter system, one has to study their influence on the 
results obtained within the Hartree approximation. 
A systematic perturbation theory for strong interactions is however lacking.  
Instead, one resorts to an expansion of diagrams to orders $1/N_c$, which we 
will investigate in this work at finite temperatures.

\section{ $1/N_c$- expansion}
\label{sec:expan}
The self-energy in the selfconsistent Hartree approximation appears of
the order ${\cal O}[1]$ as the leading term in the $1/N_c$- expansion, as can 
be seen from Eq. (\ref{selfhartree}).
In order to improve this approximation, we will study next to leading order 
diagrams, i.e. ${\cal O}[1/N_c]$- contributions. 
Therefore we make the following ans\"{a}tze for the self-energy and for the 
quark propagator
\begin{eqnarray} 
\label{12}
\Sigma(p)&=&\Sigma^H[{\rm p} ;G] + \frac{1}{N_c}\delta\Sigma[p;G]+{\cal O}
[1/N^2_c]~,\\
\label{122}
G(p)&=&G^H(p)+\frac{1}{N_c}\delta G(p)+{\cal O}(1/N^2_c)~,
\end{eqnarray} 
where the corrections to the self-energy depend in the general case on the
 full 
Green function $G(p)$ and on the 4-momentum $p$. 
The corrections to the self-energy $\delta\Sigma[p;G]$ are not yet specified
 and
will be discussed in the following section.
Using the $1/N_c$- approximation (\ref{12}) for $\Sigma[p;G]$ in the
Dyson equation (\ref{dyson}), the  $1/N_c$- expansion to the propagator
is given as
\begin{eqnarray} 
G(p)&=&\bigg(G_0^{-1}(p)-\Sigma[p;G]\bigg)^{-1}\nonumber\\
&=&\bigg(G_0^{-1}(p)-\Sigma^H[{\rm p} ;G^H]-\frac{1}{N_c}\bigg[\Sigma^H[{\rm p};
\delta G]+\delta \Sigma[p;G^H]+{\cal O}(1/N_c)\bigg]\bigg)^{-1}~.
\end{eqnarray} 
Expanding the $1/N_c$ contribution in the denominator and comparing with 
(\ref{122}),  we obtain  a selfconsistent equation for $\delta G(p)$ in the
 form
\begin{equation} 
\label{dg}
\delta G(p) = G^H(p)\Sigma^H[{\rm p} ;\delta G]G^H(p) +
G^H(p)\delta \Sigma[p;G^H]G^H(p) ~.
\end{equation} 
Note, that this consistent $1/N_c$- expansion for the quark propagator is a 
new 
result of this paper. 
In particular, the first term on the r.h.s. of (\ref{dg}) has not been 
considered in some of the previous approaches, see \cite{sandi,zhuang2}.
In order to get a closed expression, we use the fact that the functional 
dependence of the Hartree selfenergy on the $1/N_c$- corrections to the quark 
propagator $\delta G$ is known from Eq. (\ref{selfhartree}).
After insertion of $\Sigma^H[{\rm p} ;\delta G]$ on the r.h.s. of Eq.
 (\ref{dg}),
we obtain
\begin{eqnarray} 
\label{trace}
\sum_p g({\rm p} ) {\bf Tr}[\delta G(p)]&=& -K_0 N_f \sum_p g^2({\rm p} ) 
{\bf Tr} \bigg[ G^H(p) G^H(p) \bigg] \sum_k g({\rm k} ) {\bf Tr} 
[\delta G(k)]\nonumber\\
&+&\sum_p g({\rm p} ) {\bf Tr} \bigg[
G^H(p)\delta \Sigma[p;G^H] G^H(p) \bigg]\nonumber\\
&=& \frac{1}{1-J^\sigma(0)} \sum_p g({\rm p} ) {\bf Tr} 
\bigg[G^H(p)\delta \Sigma[p;G^H]G^H(p) \bigg]~,
\end{eqnarray} 
where the scalar quark loop integral $J^\sigma(0)$ is defined in Appendix 
\ref{app:int}.
The $1/N_c$- expansion of the quark condensate corresponding to that of the
propagator (\ref{122}) and the definiton of the quark condensate
(\ref{condensate}) reads
\begin{equation} 
\label{qqneu}
<{\bar q} q> = <{\bar q} q>^H + \delta <{\bar q} q> + {\cal O} [1/N_c^2]\,\,.
\end{equation} 
The $1/N_c$ correction to the condensate is obtained in closed form using the 
result (\ref{trace})
\begin{eqnarray} 
\label{delconden}
\delta <{\bar q} q>
= Z \cdot N_f \sum_p  g({\rm p} ){\bf Tr}\bigg[G^H(p) \delta \Sigma[p;G^H] 
G^H(p)\bigg]
\end{eqnarray} 
with a prefactor $Z = 1/(1-J^\sigma(0))$ as derived in (\ref{trace}) coming 
from the $1/N_c$- contributions to the Hartree self-energy 
$\Sigma^H[{\rm p},\delta G]$, see in Fig. 3.
This prefactor leads to a considerable rescaling ($Z\sim 4$) which has first 
been pointed out in Refs. \cite{volkov,lemmer} for the NJL model at zero 
temperature and is obtained here for the more general case of a nonlocal
separable interaction at finite temperature.

\section{Mesonic correlations}
\label{sec:meson}
Within the chiral quark model as defined in Section \ref{sec:model}, the 
complete set of diagrams contributing in ${\cal O} [1/N_c]$ to the self-energy 
is given in Fig. 4.
The double line corresponds to the RPA - type partial resummation of the chain 
of bubble diagrams, where the quark - antiquark loop in Hartree approximation 
defines the polarization functions $J^{\phi }(p-k)$ in the scalar and
pseudoscalar channel ($\phi  = \sigma,\pi$), see Appendix \ref{app:int}.

The ${\cal O} [1/N_c]$ self-energy contribution is given by
\begin{eqnarray} 
\label{dsm}
\delta\Sigma[p;G^H] =  K_0 \sum_k g^2\bigg(\frac{|{\bf p+k}|}{2}\bigg)
\bigg[\frac{G^H(k)}{1-J^\sigma(p-k)}-
(N_f^2-1)\frac{\gamma_5G^H(k)\gamma_5}{1-J^\pi(p-k)}\bigg]~.
\end{eqnarray} 
The denominators $1 - J^{\phi }(p-k)$ occur due to the resummation and thus 
strong correlations can be described.
Note that the $1/N_c$ self-energy is a dynamical quantity and has not yet
been solved in its complexity.
The most dramatic effect is the occurence of collective excitations in the
quark - antiquark channel when ${\rm Re} J^\phi  (P = M_\phi ) = 1$ (and
${\rm Im} J^\phi (P = M_\phi ) = 0$) which correspond to mesonic bound states.
In what follows we restrict us to the consideration of bound states only
and use an expansion of the polarization function at the mesonic poles
(pole approximation) which leads to the introduction of meson propagators
and meson- quark-antiquark form factors $g_{\phi  q{\bar q}}$ 
(see Appendix \ref{app:int})
\begin{equation} 
\label{mesprop}
\frac{1}{1-J^{\phi }(P)} \approx \frac{1}{M_{\phi }^2-P^2}\frac{g^{2}_{\phi  q{\bar q}}
(M_{\phi })}{N_fK_0}\,.
\end{equation} 
The full treatment of the RPA approximation which contains bound and
 scattering 
states is possible for the separable interaction and will be regarded in an 
additional work.

Using the expression  (\ref{dsm}, \ref{mesprop}) for the self-energy and the
short notation with  $\Gamma^\sigma = 1_D$ and $\Gamma^\pi=i\gamma_5$,
we obtain
\small
\begin{eqnarray} 
\label{cond2}
\delta<{\bar q} q>^\phi &=&\frac{g^2_{\phi {\bar q} q}}{1-J^\sigma(0)}
\int\frac{d^3{\bf p}}{(2\pi)^3}g({\rm p} )\int\frac{d^3{\bf k}}{(2\pi)^3}g^2
\bigg(\frac{|{\bf p + k}|}{2}\bigg)\nonumber\\
& &\times\int\frac{dp_0}{2\pi}\int\frac{dk_0}{2\pi}\frac{1}{M^2_\phi -(k-p)^2}
{\bf Tr}\bigg[G^H(p)\Gamma^\phi  G^H(k)\Gamma^\phi  G^H(p)\bigg]~.
\end{eqnarray} 
\normalsize
After evaluation of the Dirac trace
\begin{eqnarray} 
& &\frac{1}{M^2_\phi -(k-p)^2}
{\bf Tr}\bigg[G^H(p)\Gamma^\phi  G^H(k)\Gamma^\phi  G^H(p)\bigg]=\nonumber \\
& &-4\bigg(\frac{m({\rm p} )\pm m({\rm k} )}{(p^2-m^2({\rm p} ))
(k^2-m^2({\rm k} ))[(k-p)^2-M^2_\phi ]}+m({\rm p} )
\bigg[\frac{1}{(p^2-m^2({\rm p} ))^2[(k-p)^2-M^2_\phi ]}\nonumber \\
& &-\frac{M^2_\phi -(m({\rm p} )\pm m({\rm k} ))^2}
{(p^2-m^2({\rm p} ))^2(k^2-m^2({\rm k} ))[(k-p)^2-M^2_\phi ]}
-\frac{1}{(p^2-m^2({\rm p} ))^2(k^2-m^2({\rm k} ))}\bigg]\bigg)~,
\end{eqnarray} 
we perform the Matsubara summation (see Appendix \ref{app:matsum}) and 
obtain as the final result for the $1/N_c$ mesonic contributions (\ref{cond2}) 
to the quark condensate
\small
\begin{eqnarray} 
\label{condT}
\delta<{\bar q} q>^\phi &=& \frac{g^2_{\phi {\bar q} q}}{1-J^\sigma(0)}
\int\frac{d^3{\bf p}}{(2\pi)^3} g({\rm p} )
\int\frac{d^3{\bf k}}{(2\pi)^3} g^2\bigg(\frac{|{\bf p+k}|}{2}\bigg)\nonumber\\
& &\times 
\bigg\{\frac{2 m^H({\rm p} )}{E^2({\rm p} )}
\bigg(\frac{1-2f(E({\rm p} ))}{2 E({\rm p} )}-\frac{f(E({\rm p} ))
[1-f(E({\rm p} ))] }{T}\bigg)\nonumber\\
& &\times
\bigg(\frac{1-2f(E({\rm k} ))}{2 E({\rm k} )}
-\frac{1+2n(E_\phi ({\bf k-p}))}{2 E_\phi ({\bf k-p})}\bigg)\nonumber \\
& &+
\bigg(\bigg[\frac{ [1-f(E({\rm p} ))-f(E({\rm k} ))][1+n(E({\rm p} )
+E({\rm k} ))+n(E_\phi ({\bf k-p}))]} 
{E^3({\rm p} )E({\rm k} )E_\phi ({\bf k-p})
(E_\phi ({\bf k-p})+E({\rm p} )+E({\rm k} ))}\nonumber \\
& &\times
\bigg\{ E^2({\rm p} ) [m^H({\rm p} )\pm m^H({\rm k} )] + m^H({\rm p} )
[M^2_\phi -[m^H({\rm p} ) \pm m^H({\rm k} )]^2]\\
&\times&\left[ \frac{E_\phi ({\bf k-p})+2 E({\rm p} )+E({\rm k} )}
{E_\phi ({\bf k-p})+E({\rm p} )+E({\rm k} )} 
+\frac{E({\rm p} ) f(E({\rm p} ))[1-f(E({\rm p} ))] }{T} \right] \bigg\}
\nonumber\\
& &
+[E({\rm k})\rightarrow -E({\rm k})]\bigg]
+[E({\rm p})\rightarrow -E({\rm p})]\bigg)\bigg\}\nonumber  ~,
\end{eqnarray} 
\normalsize
with the energies $E_\phi ({\bf k-p})=\sqrt{({\bf k-p})^2+M_\phi ^2}$ and the
bosonic distribution function $n (E) = \left[\mbox{exp}
\left(E/T \right)-1 \right]^{-1}$. 
The upper sign holds for the scalar, the lower one for the pseudoscalar
meson, respectively. 
The  ${\cal O}[1/N_c]$ contribution (\ref{condT}) consists of two parts,
and the numerical analysis shows that the contribution due to mesonic 
correlations is dominated by the first one, i.e.
\small
\begin{eqnarray} 
\label{condT1}
\delta<{\bar q} q>^\phi  &\approx& \frac{g^2_{\phi {\bar q} q}}{1-J^\sigma(0)}
\int\frac{d^3{\bf p}}{(2\pi)^3} g({\rm p} )
\int\frac{d^3{\bf k}}{(2\pi)^3} g^2\bigg(\frac{|{\bf p + k}|}{2}\bigg)
\frac{m^H({\rm p} )[1-2f(E({\rm p} ))]}{E^3({\rm p} )}\nonumber\\
& &\times
\bigg(\frac{1-2f(E({\rm k} ))}{2 E({\rm k} )}-\frac{1+2n(E_\phi ({\bf k-p}))}
{2 E_\phi ({\bf k-p})}\bigg)~,
\end{eqnarray} 
\normalsize
which has a simpler structure than (\ref{condT}).

In order to compare our results with previous works we will now discuss the 
$T = 0$ case.
At zero temperature the Fermi and Bose distribution functions vanish.
In the $T = 0$ limit of Eqs. (\ref{condTH}) and (\ref{condT}) we obtain
\small
\begin{eqnarray} 
\label{condT0}
\delta<{\bar q} q>^\phi &=&\frac{g^2_{\phi {\bar q} q}}{1-J^\sigma(0)}
\int\frac{d^3{\bf p}}{(2\pi)^3}g({\rm p} )
\int\frac{d^3{\bf k}}{(2\pi)^3}g^2\bigg(\frac{|{\bf p + k}|}{2}\bigg)
\bigg(\frac{ m^H({\rm p} )}{2 E^3({\rm p} )}
\cdot\bigg(\frac{1}{E({\rm k} )}-\frac{1}{E_\phi ({\bf k-p})}\bigg)\nonumber \\
&+&\frac{[M^2_\phi -[m^H({\rm p} )\pm m^H({\rm k} )]^2]
[E_\phi ({\bf k-p})+2 E({\rm p} )+E({\rm k} )]m^H({\rm p} )}
 {E^3({\rm p} )E({\rm k} )E_\phi ({\bf k-p})(E_\phi ({\bf k-p})+E({\rm p} )
+E({\rm k} ))^2}\bigg)\nonumber \\
&+&\frac{ [m^H({\rm p} )\pm m^H({\rm k} )] }
{E({\rm p} )E({\rm k} )E_\phi ({\bf k-p})(E_\phi ({\bf k-p})+E({\rm p} )
+E({\rm k} ))}\bigg)~.
\end{eqnarray} 
\normalsize
In the following section we present the numerical evaluation and discussion 
of the  above  $1/N_c$- corrections to the quark condensate.
\section{Numerical results and discussion}
\label{sec:disc}
In section \ref{sec:model} we have introduced a general nonlocal interaction
kernel in separable form.
In order to compare the numerical results with previous
approaches within the NJL model, we will restrict
us in this paper to the discussion of a cut-off formfactor
\begin{eqnarray} 
\label{cutoff}
g\bigg(\frac{|{\bf p + k}|}{2}\bigg) &=& \Theta\bigg(1 -
\frac{|{\bf p + k}|}{2 \Lambda_{\rm NJL}}\bigg)~.
\end{eqnarray} 
The chiral quark model with soft formfactors as, e.g. a Gaussian one,
has been discussed in Refs. \cite{short,physref}.

After fixing the parameters of
the model as described in Appendix \ref{app:int},
we obtain for the quark condensate
in the Hartree approxim ation $<{\bar u}u>^H=-(250 $~MeV$)^3$ and for
the quark mass
$m^H = 300$ MeV
in agreement with the well-known data of the literature
\cite{nJl,njl,NJL,hatsuda}.

In the next step, discussed in Section \ref{sec:meson}, we have included
dynamical selfenergy contributions due to mesonic correlations.
Compared with the Hartree term (\ref{condTH}) where we have to solve
a one-loop integral, the next to leading order contributions are two-loop 
integrals which after summation over both Matsubara frequencies
($k_0,p_0$) reduce to three-dimensional integrals over the
variables ${\rm k} ,{\rm p} ,z$ , where
$z = \cos\theta$, if $\theta$ denotes the angle between the momenta
${\bf k}$ and ${\bf p}$.

At first we want to discuss the $T = 0$ limit.
An open question which occurs in the conventional NJL model is the choice of 
the cut-off for the second momentum integral in Eqs. (\ref{condT}) and 
(\ref{condT0}) over ${\rm k} $.
A very crude approximation presented in Ref.\cite{zhuang2} is the omission of 
the second integral by assuming that ${\bf k} = 0$.
In Refs. \cite{volkov,lemmer} the additional cut-off ${\bar \Lambda}$ was 
discussed.
Ref.\cite{volkov} assumes that ${\bar \Lambda} = \Lambda_{\rm NJL}$ and
in Ref.\cite{lemmer} upper and lower limits are determined from a calculation
of $f_\pi$ in ${\cal O} [1/$N$_c]$.
In the formulation we have chosen such a problem does not exist since the 
integrals are regularized in the separable approach by the proper treatment of 
the formfactors.
The parameter $\Lambda_{\rm NJL}$ in the cut-off formfactor (\ref{cutoff})
regularizes the integral over ${\rm p} $. 
The upper limit of the ${\rm k} $- integration is given by 
${\bar \Lambda} = -pz+\sqrt{4\Lambda^2_{\rm NJL}-p^2(1-z^2)}$  
and runs between $\Lambda_{\rm NJL} < {\bar \Lambda}<3\Lambda_{\rm NJL}$.
Note that in solving (\ref{condT}) one has to check the integral limits for
each term separately due to different  combinations of formfactors
partly hidden in the momentum dependent quark mass (\ref{mh}).
Thus we have removed the ambiguity in regularizing the second momentum 
integration which occured in the previous approaches to the $1/N_c$- expansion 
in the NJL model.

The result for such a calculation in the $T = 0$ limit  is that for fixed model
parameters the absolute value of the condensate is decreased by 20\% compared 
to the Hartree- approximation. 
For comparison, a decrease of the quark mass due to the mesonic correlations in
$1/N_c$ at $T = 0$ has been obtained in Ref.\cite{volkov}.
This result can be understood qualitatively since the Hartree- contribution 
(\ref{condTH}) and the 1/N$_c$ mesonic contribution (\ref{condT}) to the quark 
condensate have opposite sign, the latter one being smaller in magnitude.
 
Let us consider the finite temperature case. 
In order to compare the temperature behaviour of the quark condensate in both 
models (Hartree approximation and Hartree approximation with mesonic
correlations) we have to fix the parameters such that the same values for the 
observables at $T = 0$ are obtained, see Appendix A. 
The numerical evaluation of the final result for the quark condensate  
is shown in Fig. 5. 
Paying attention to the shape of the chiral phase transition,
we observe that the inclusion of $1/N_c$ mesonic correlations shifts 
the chiral symmetry restoration to lower temperatures when compared with the 
simple Hartree approximation.
This finding is mainly due to the smaller $q \bar q$ coupling constant $K_0$ 
for the model with mesonic correlations.
 
\section{Conclusions}
\label{sec:end}
In conclusion we have obtained the following new results:
(i) a consistent $1/N_c$- expansion for the self-energy as well as
for the propagator and a closed formula for
the quark condensate up to the order  $1/N_c$,
(ii) a finite temperature result for the $1/N_c$ quark condensate within 
the Matsubara formalism,
(iii) a consistent regularization of the two loop diagrams.

The numerical evaluation for a NJL-type model shows that compared with the 
Hartree approximation mesonic correlations lead to a decrease of the absolute 
value of the quark condensate at $T = 0$. After having compensated this effect 
of quantum fluctuations at $T = 0$ by readjusting the model parameters 
($\Lambda, m_0, K_0$), the account for thermaly excited mesonic correlations
shifts the onset of the chiral symmetry restoration to lower 
temperatures in comparison to the Hartree approximation. 
This behaviour seems to be closer to the recent results of lattice 
calculations where the condensate remains unchanged with temperature up to 
the chiral transition which occurs at $T\approx 150$ MeV for $N_f=2$ 
\cite{lattice}.

The inclusion of quark-antiquark correlations is of principal interest because 
the treatment of the medium in free quasiparticle approximation seems not 
to be appropriate at low temperatures.
In contrast, in this region the mesonic degrees of freedom are expected 
to be relevant.
This is supported by the fact that in the low temperature limit 
(T $ \stackrel{<}{\sim} 50 $ MeV)  also other thermodynamic properties of a 
quark-meson system (e.g. the pressure) are dominated by mesonic contributions 
\cite{zhuang}.
The presented $1/N_c$- expansion should be considered as a first step in 
including mesonic correlations. 
However, at temperatures where the chiral phase transition occurs, higher 
orders of the $1/N_c$- expansion may become important.

Within the present approach, the treatment of the two-particle correlations
was given in the usual pole approximation (\ref{mesprop}) for the 
$q \bar q $ T-matrix.
A next step in the evaluation of quark-antiquark correlations is the 
inclusion of the contribution of scattering states which will be 
considered in a forthcoming paper.
In this way the account of the corrections due to two-particle 
correlations will be completed on the basis of the approach presented here.

\section*{Acknowledgement}
Our thanks go to the {\sc Deutsche Forschungsgemeinschaft} which gave a
 research 
grant to one of us (Yu. K.).
G.R. and S.S. are grateful for the hospitality  during a stay in Dubna. 
M.K.V. acknowledges the support provided  by INTAS grant No. W 94-2915 and 
RFFI grant No. W 93-02-14411 as well as the hospitality extended to him during 
a stay in Rostock.

\begin{appendix}

\section{Table of integrals and parameter fixing}
\label{app:int}

The polarization operators introduced in Eq. (\ref{dsm}) of the main text are 
defined as
\begin{equation} 
J^{\phi }(P)=-K_0N_f\sum_q g^2({\rm q} ){\bf Tr}\bigg[\Gamma^\phi G^H(q+P/2)
\Gamma^\phi  G^H(q-P/2)\bigg]\,\,.
\end{equation} 
The temperature dependent meson masses are obtained from the solution of the 
Bethe-Salpeter equation
\begin{eqnarray}
\label{bse}
1-J^{\phi }(P_0=M_{\phi }(T),{\bf P}=0)  =0 ~~,
\end{eqnarray}
where the polarization operators  $J^{\phi }(P_0)$ after evaluation of the 
Dirac trace and angular integrationare given by
\begin{eqnarray}
\label{jpion}
J^\pi(P_0)&=& \frac{2 K_0 N_f}{\pi^2} \int d{\rm q}{\rm q}^2 {g^2({\rm q} )}~
\frac{E({\rm q})}{E^2({\rm q}) - (P_0 / 2)^2 } ~[1-2 f (E({\rm q} ))]~~ , \\
\label{jsigma}
J^\sigma(P_0)&=& \frac{2 K_0 N_f}{\pi^2}
\int d{\rm q}  {\rm q} ^2 \frac{g^2({\rm q} )}{E({\rm q} )}~~
\frac{{\rm q} ^2}{E^2({\rm q} ) - (P_0/ 2)^2 } \,[1-2 f (E({\rm q} ))]~~,   
\label{sigmamass}
\end{eqnarray}
The quark meson coupling constants introduced in Eq. (\ref{mesprop}) are 
evaluated in the rest frame of the $q \bar q$ pair (${\bf P} = 0$), where
\begin{equation} 
g^{-2}_{\phi  q{\bar q}}(M_{\phi })= \frac{1}{2K_0N_f}
\frac{d}{dP^2}J^{\phi }(P)\bigg|_{P^2=M^2_{\phi }}
\approx\frac{1}{2K_0N_f}\frac{d}{2P_0dP_0}J^{\phi}(P_0,0)\bigg|_{P_0=M_\phi}~.
\end{equation} 
Using Eqs. (\ref{jpion}) and (\ref{jsigma}), they are given by the following 
integrals
\begin{eqnarray}
\label{gqq}
g_{\pi q{\bar q}}^{-2}(M_\pi)
& = & \frac{1}{2 \pi^2 } \int     {d{{\rm q} }} ~~ {\rm q} ^2 \,\,
g^2({\rm q} ) \,\, \frac{E({\rm q} )}{(E^2({\rm q} )-M_\pi^2/4)^2}
\biggl[ 1-2\,f( E({\rm q}) )\biggr]~~,\\
g_{\sigma q{\bar q}}^{-2}(M_\sigma)
& = & \frac{1}{2 \pi^2  } \int     {d{{\rm q} }} ~~ {\rm q}^2 ~~
g^2({\rm q})~~\frac{{\rm q}^2}{E({\rm q})}~\frac{1}{(E^2({\rm q})-
M_\sigma ^2/4)^2}
\biggl[ 1-2\,f( E({\rm q} ) )\biggr]~~.    \nonumber
\end{eqnarray}
The pion decay constant which we use for the parameter fixing at zero 
temperature is calculated by 
\cite{physref}
\begin{eqnarray} 
\label{fpi}
f_\pi& =& \frac{\sqrt{N_c}g_{\pi q \bar q}}{2 \pi^2} \int d{\rm q}~~{\rm q}^2
 ~~
g({\rm q} )~~\frac{m^H({\rm q} )}{E({\rm q} )(E^2({\rm q} )-M_\pi^2/4)}~.
\end{eqnarray} 
The model contains three parameters: the coupling constant $K_0$, the current 
quark mass $m_0$ and the range of the formfactor of the potential.
We fix these 3 parameters to reproduce the pion mass ($M_\pi = 140$ MeV) 
Eqs.(\ref{bse}) and (\ref{jpion}), the pion decay constant ($f_\pi = 93$ MeV) 
Eqs. (\ref{gqq}) and (\ref{fpi}) and the quark condensate 
($-<q \bar q>^{1/3}~= 250$ MeV) at zero temperature.
The resulting parameter sets are given in Table 1. for both the Hartree 
approximation with and without the $1/N_c$ contribution from mesonic 
correlations, respectively. 
The parameters for the Hartree approximation are similar to those of the
standard NJL- model, see Refs. \cite{njl,NJL,sandi,volkov,lemmer,zhuang2}.

\section{Matsubara-Summation}
\label{app:matsum}
 
The following formulae summarize the results of the one and two loop 
Matsubara-sums performed in this paper:
\begin{equation} 
\int \frac{dk_0}{2\pi}\frac{1}{k^2_0-E^2({\rm k} )}=
-\frac{1-2f(E({\rm k} ))}{2E({\rm k} )}~,
\end{equation} 
\begin{eqnarray} 
\int \frac{dp_0}{2\pi}\frac{1}{(p^2_0-E^2({\rm p} ))^2}&=&
\frac{1}{4E^2({\rm p} )}\bigg(\frac{1-2f(E({\rm p} ))}{2E({\rm p} )}
-\frac{f(E({\rm p} ))(1-f(E({\rm p} )))}{T}
\nonumber\\ \nonumber\\
& &
+\{E({\rm p} )\rightarrow -E({\rm p} )\}\bigg)~,
\end{eqnarray} 
\begin{equation} 
\int \frac{dk_0}{2\pi}\frac{1}{(k_0-p_0)^2-E^2_\phi ({\bf k-p})}=
-\frac{1+2n(E_\phi ({\bf k-p}))}{2E_\phi ({\bf k-p})}~,
\end{equation} 
\begin{eqnarray} 
\int \frac{dk_0}{2\pi}& &
\frac{1}{((k_0-p_0)^2-E^2_\phi ({\bf {\rm k} -{\rm p} }))
(k^2_0-E^2({\rm k} ))}=\nonumber\\
\nonumber\\
& & 
\bigg(\frac{f(E({\rm k} ))+n(E_\phi ({\bf k-p}))}{4(p_0+E_\phi ({\bf k-p})
-E({\rm k} ))E({\rm k} )E_\phi ({\bf k-p})}\nonumber \\
& &+\{E({\rm k} )\rightarrow -E({\rm k} )\}\bigg)+
\{E_\phi ({\bf k-p})\rightarrow -E_\phi ({\bf k-p})\}~,
\end{eqnarray} 
\begin{eqnarray} 
\int \frac{dp_0}{2\pi}& &\frac{1}{(p^2_0-E^2({\rm p} ))
(p_0+E_\phi ({\bf k-p})-E({\rm k} ))}=\nonumber\\
\nonumber\\
& &
\frac{f(E({\rm p} ))-f(E({\rm k} ))}{2E({\rm p} )
(E_\phi ({\bf k-p})-E({\rm k} )+E({\rm p} ))}+
\{E({\rm p} )\rightarrow -E({\rm p} )\}~.
\end{eqnarray} 
\end{appendix}


\newpage
\noindent
{\Large \bf Table caption}\vspace{2cm}\\

\normalsize
\parindent 0cm
\noindent
{\bf Table 1.}
Sets of parameters for the Hartree approximation without and with 
the inclusion of mesonic correlations for a fixing scheme described in the 
text.
\\

\newpage
\begin{table}[hbt]
\begin{center}
\begin{tabular}{|c|c|c|c|}
\hline
Approximations&$\Lambda$[MeV]&$m_0$[MeV]&$K_0$[MeV$^{-2}$]\\ \hline 
Hartree&660&5.35&9.45\\ \hline
Hartree+mesonic correlations&765&4.4&6.49\\ \hline
\end{tabular}
\vspace*{1cm}
\caption{
\label{parameters}}
\end{center}
\end{table}

\newpage
\noindent

{\Large \bf Figure captions}\vspace{2cm}\\

\normalsize
\parindent 0cm
\noindent
{\bf Figure 1.}
The Dyson equation with the full self-energy.\\

\noindent
{\bf Figure 2}.
The Dyson equation in selfconsistent Hartree approximation.\\

\noindent
{\bf Figure  3}.
$1/N_c$-expansion of the quark condensate.                 \\

\noindent
{\bf Figure 4}.
$1/N_c$- approximation for the self-energy. The scalar- and pseudoscalar
correlations are described by a RPA- type partial summation of bubble 
diagrams. \\

\noindent
{\bf Figure 5}.   The quark condensate as a function of the temperature in 
selfconsistent Hartree approximation (dotted line) and with inclusion of
mesonic correlations (solid line).

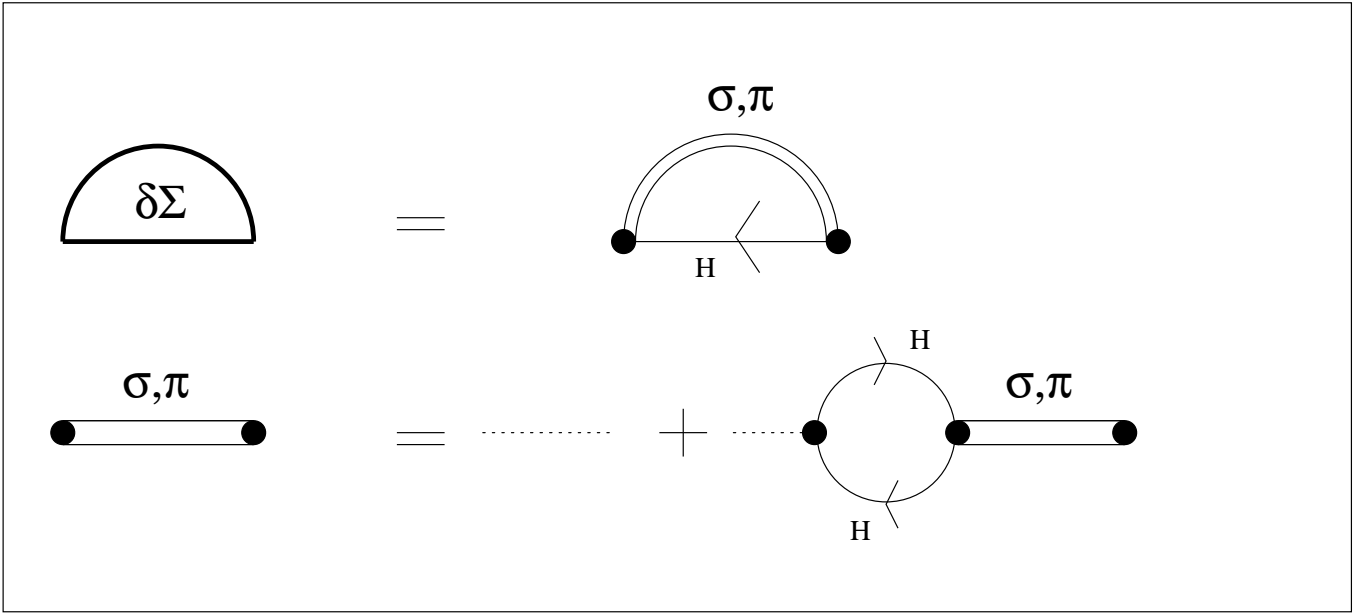
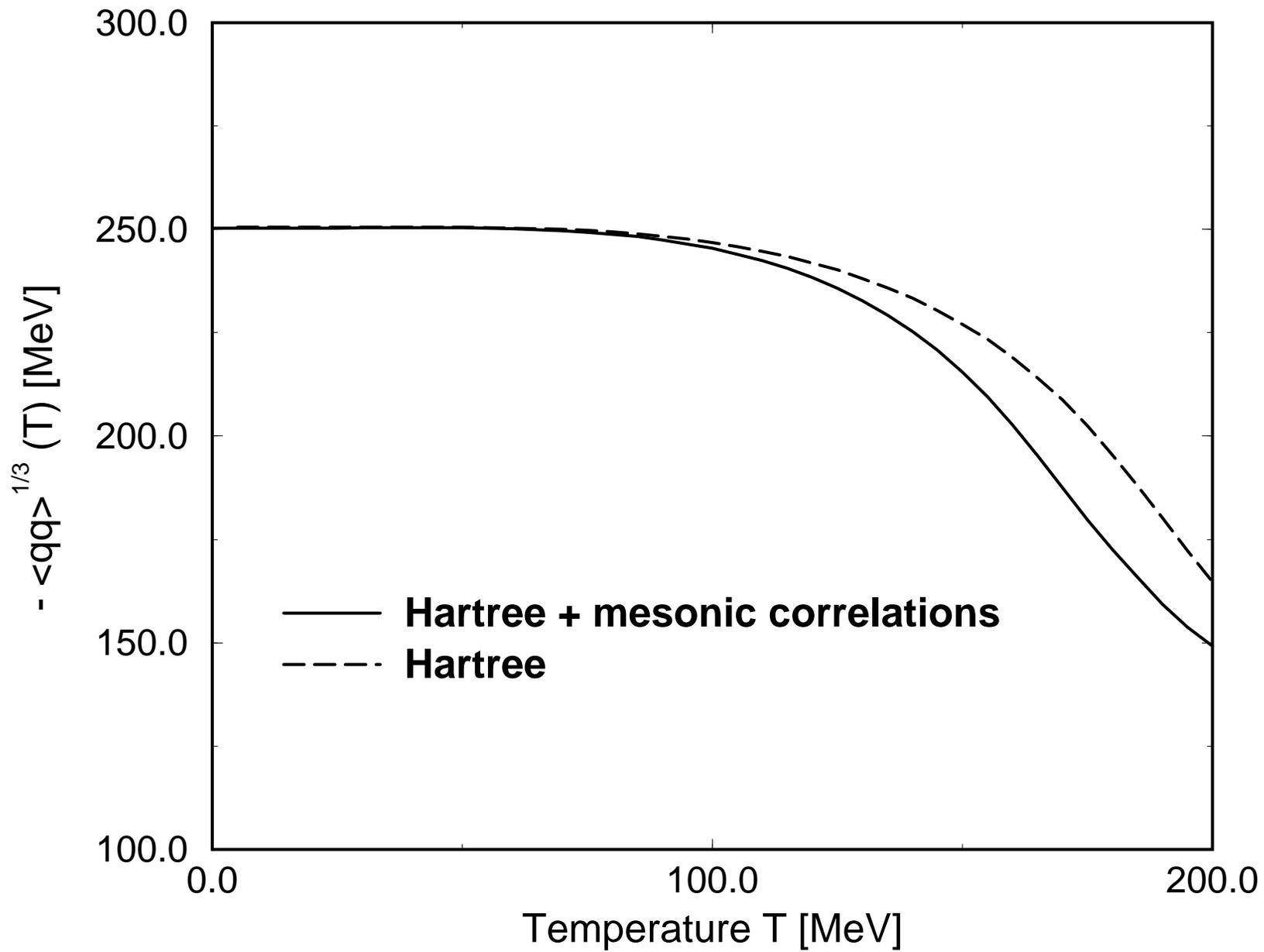
\end{document}